\documentclass[]{aa} % for the letters 
\usepackage{txfonts,epsfig,graphicx,natbib,url,twoopt,amsmath,algorithm2e}
\usepackage{color}

\newcommand{\mubold}{\mbox{\boldmath$\mu$}}

\newcommand{\epsilonbold}{\mbox{\boldmath$\epsilon$}}

\newcommand{\Sigmabold}{\mbox{\boldmath$\Sigma$}}
\newcommand{\sigmabold}{\mbox{\boldmath$\sigma$}}

\newcommand{\alphabold}{\mbox{\boldmath$\alpha$}}

\usepackage{natbib,twoopt}
\bibpunct{(}{)}{;}{a}{}{,}             %% natbib format for A&A and ApJ
\makeatletter
  \newcommandtwoopt{\citeads}[3][][]{\href{http://adsabs.harvard.edu/abs/#3}%
    {\def\hyper@linkstart##1##2{}%
     \let\hyper@linkend\@empty\citealp[#1][#2]{#3}}}
  \newcommandtwoopt{\citepads}[3][][]{\href{http://adsabs.harvard.edu/abs/#3}%
    {\def\hyper@linkstart##1##2{}%
     \let\hyper@linkend\@empty\citep[#1][#2]{#3}}}
  \newcommandtwoopt{\citetads}[3][][]{\href{http://adsabs.harvard.edu/abs/#3}%
    {\def\hyper@linkstart##1##2{}%
     \let\hyper@linkend\@empty\citet[#1][#2]{#3}}}
  \newcommandtwoopt{\citeyearads}[3][][]%
    {\href{http://adsabs.harvard.edu/abs/#3}
    {\def\hyper@linkstart##1##2{}%
     \let\hyper@linkend\@empty\citeyear[#1][#2]{#3}}}
\makeatother

\begin{document}

\title{Bayesian least squares deconvolution}

\author{A. Asensio Ramos\inst{1,2} \and P. Petit\inst{3,4}}

\institute{
 Instituto de Astrof\'{\i}sica de Canarias, 38205, La Laguna, Tenerife, Spain; \email{aasensio@iac.es}
\and
Departamento de Astrof\'{\i}sica, Universidad de La Laguna, E-38205 La Laguna, Tenerife, Spain
\and
Universit\'e de Toulouse, UPS-OMP, Institut de Recherche en Astrophysique et Plan\'etologie, Toulouse, France 
\and
CNRS, Institut de Recherche en Astrophysique et Plan\'etologie, 14 Avenue Edouard Belin, F-31400 Toulouse, France
}
             
  \date{Received ---; accepted ---} 

  \abstract
  {}
  {To develop a fully Bayesian least squares deconvolution (LSD) that can be applied to the reliable detection of magnetic
  signals in noise-limited stellar spectropolarimetric observations using multiline techniques.}
  {We consider LSD under the Bayesian framework and we introduce a flexible Gaussian Process (GP) prior for the LSD profile. This prior 
  allows the result to automatically adapt to the presence of signal. We exploit several linear algebra identities to
  accelerate the calculations. The final algorithm can deal with thousands of spectral lines in a few seconds.}
  {We demonstrate the reliability of the method with synthetic experiments and we apply it to real spectropolarimetric observations of magnetic
  stars. We are able to recover the magnetic signals using a small number of spectral lines, together with the uncertainty at each
  velocity bin. This allows the user to consider if the detected signal is reliable. The code to compute the Bayesian LSD
  profile is freely available.}
  {}

   \authorrunning{Asensio Ramos \& Petit}
   \titlerunning{Bayesian LSD}
   \maketitle
%
%________________________________________________________________

\section{Introduction}
The phenomena taking place on the atmospheres of solar-type stars is controlled by
magnetic fields. One of the most straightforward ways of detecting and measuring
such magnetic fields is through spectropolarimetry. The observation of the intensity (Stokes $I$), linear
polarization (Stokes $Q$ and $U$) and circular polarization (Stokes $V$) allows us to
remotely infer the strength and orientation of magnetic fields. 

During the last decades, we have been able to detect magnetic fields in cool stars.
The first efforts \citep{donati92} detected the presence of polarimetric signals in the most active of them. 
In these cases, it was necessary to reach a very high polarimetric sensitivity to detect 
these signals in individual spectral lines. In general, tough, the polarization level in almost all cool
stars is below our detection limits and we require post-processing techniques to enhance the detection 
rate. The first solution to this problem was the development of multiline techniques \citep{semel89,semel96}, who
averaged many photospheric spectral lines to improve the effective polarimetric sensitivity. This technique
works under the assumption that noise is uncorrelated and adds up to zero when many spectral lines are
summed. Meanwhile, the spectral line itself presents correlations that will make it appear above the noise
after adding many spectral lines. This idea was later improved with the introduction of the least-squares
deconvolution technique \citep[LSD;][]{donati97}, which has facilitated the detection of polarimetric
signals in a wide variety of stars and the generation of a set of interesting surveys for the detection
of magnetic fields in different stellar types \citep[e.g.,][]{wade12,marsden14,martins15}. 

Recently, the LSD method has been critically analyzed by \cite{kochukhov10}, pointing out the difficulties
on the interpretation of the final LSD profile in terms of a normal spectral line. In the recent years, 
several variations of the LSD method have been presented. These methods try to overcome some of the fundamental
problems with LSD. For instance, \cite{sennhauser09} suggested to introduce an improvement to nonlinearly
deal with the presence of blends in spectral lines, much close to what really happens during the formation
of the spectral line. \cite{carroll08} and \cite{marian_pcafilter08} generalized LSD to account for the deviations from the
assumption of a single common profile for all spectral lines. This was done by using the principal 
component analysis (PCA) decomposition of all the considered lines and truncating the decomposition to get rid
of the noise. \cite{semel09} introduced the idea of multiline Zeeman signatures as detectors of the 
presence of magnetic fields. Finally, \cite{kochukhov10} introduced multiprofile LSD, a generalization of LSD
which models each spectral line as the linear addition of several profiles, much in the line of the PCA approach.
These authors also discuss a simple modification of LSD to regularize the solution using a first-order Tikhonov
regularization, which favors smooth solutions.

In this paper we treat the LSD from a Bayesian perspective and consider the use of a flexible
prior for the expected LSD profile. This introduces a regularization that automatically adapts
to the data. The method is fast and competitive in computing time with LSD. One of the main
advantages is that, given that the method we propose is fully Bayesian, we are able to give
sensible error bars to the detection of LSD signatures.

\section{Bayesian LSD}
The LSD technique is based on assuming that the circular polarization profiles of
a set of selected spectral lines from a star correspond to a common basic Zeeman
signature. The spectral shape of this signature does not change and only
the amplitude is modified from line to line. The validity of this approximation is based on two 
assumptions. First, the line has to be in the so-called weak field regime, which happens when the magnetic field is sufficiently
weak so that the Zeeman splitting is small as compared with the Doppler width of the spectral line. 
Second, the line has to be weak, so that no saturation effects are detected.
Under these simplifying assumptions, the radiative transfer equation can be solved analytically and it can be demonstrated that the 
Stokes $V$ profile is proportional to the wavelength derivative of the intensity profile \citep[e.g.,][]{landi_landolfi04}. 
In this case, \cite{donati97} demonstrated that it is possible to write the Stokes $V$ profile for line $i$ as:
\begin{equation}
V_i(v) = \alpha_i Z(v),
\end{equation}
where $v$ is the displacement with respect to the central wavelength of the line in velocity units
and $\alpha_i$ is a proportionality constant that depends on the properties of the line. \cite{donati97}
also showed that these constants are $\alpha_i=\lambda_i d_i g_i$, with $\lambda_i$ the central wavelength of the 
line, $d_i$ is the depth of the line, and $g_i$ is the effective Land\'e factor of the line.

With a model for all the lines at hand, we can write the generative model. It states
how a given observed spectral line is modeled in terms of some parameters. Let us assume that
we observe $N_\mathrm{lines}$ spectral lines, each one sampled at $N_v$ velocity positions. Then,
we model our measurements with:
\begin{equation}
V_i(v_j) = \alpha_i Z(v_j) + \epsilon_{ij}.
\end{equation}
The previous expression indicates that our measured Stokes $V$ profile at a given
velocity position and for a certain spectral line is just a weight times the 
common signature at this very same velocity with some added noise, $\epsilon_{ij}$. This noise
can potentially be different for all velocity positions of all the spectral lines and even accommodate
the more general case in which there is non-zero correlation between different pixels.
Note that the generative model 
can be written in matrix notation as:
\begin{equation}
\mathbf{V} = \mathbf{W} \mathbf{Z} + \epsilonbold
\end{equation}
where the matrix $\mathbf{W}$ is given by:
\begin{equation}
\mathbf{W} = \left[
\begin{array}{ccccc}
\alpha_1 & 0 & 0 & 0 & \cdots \\
\alpha_2 & 0 & 0 & 0 & \cdots \\
\alpha_3 & 0 & 0 & 0 & \cdots \\
\cdots & \cdots & \cdots & \cdots & \cdots\\
0 & \alpha_1 & 0 & 0 & \cdots \\
0 & \alpha_2 & 0 & 0 & \cdots \\
0 & \alpha_3 & 0 & 0 & \cdots \\
\cdots & \cdots & \cdots & \cdots & \cdots\\
\end{array}
\right]
\end{equation}
In other words, the matrix elements are given by $W_{ij}=\alpha_j \delta_{j,i \bmod j}$.

\subsection{Standard LSD}
It is very simple to obtain the standard LSD method in a Bayesian framework. To this end, 
let the observations $\mathbf{V}$ represent 
the Stokes $V$ profiles of line $i$ at wavelength $j$ ordered as $\mathbf{V} = [V_1(v_1),V_2(v_1),V_3(v_1),\ldots,V_1(v_2),V_2(v_2),\ldots]^T$. 
Likewise, we define the $\mathbf{Z}$ vector as the sampled common signature, given by $\mathbf{Z}=[Z(v_1),Z(v_2),\ldots]^T$.
With these definitions, we can apply
the Bayes theorem to obtain the probability distribution for the common signature given all the
observed data:
\begin{equation}
p(\mathbf{Z}|\mathbf{V}) = \frac{p(\mathbf{V}|\mathbf{Z})
p(\mathbf{Z})}{p(\mathbf{V})}
\end{equation}
where $p(\mathbf{Z}|\mathbf{V})$ is the posterior distribution for the common signature given the observations,
$p(\mathbf{V}|\mathbf{Z})$ is the likelihood or sampling distribution (that gives the probability of 
getting a set of observed spectral lines conditioned on a common signature), $p(\mathbf{Z})$ is the prior
distribution for the common signature (that encodes all the a-priori information we have for $Z(v)$).
Finally, $p(\mathbf{V})$ is the marginal likelihood or evidence that normalizes the posterior to make
it a proper probability distribution. Given that the marginal likelihood does not depend on $\mathbf{Z}$,
it is usual to drop it.

Even though the equations are valid for the general correlated case, we assume for simplicity
that the noise is uncorrelated and with variance $\sigma_{ij}^2$ for line $i$ at velocity point $j$.
Under this assumption, the likelihood can be written as:
\begin{equation}
p(\mathbf{V}|\mathbf{Z}) = \mathcal{N}(\mathbf{V}|\mathbf{W} \mathbf{Z}, \mathrm{diag}(\sigmabold^2))
\label{eq:likelihood}
\end{equation}
where $\mathcal{N}(\mathbf{x}|\mubold,\mathbf{C})$ represents a multivariate Gaussian
distribution for variable $\mathbf{x}$ with mean vector $\mubold$ and covariance
matrix $\mathbf{C}$. The vector $\sigmabold^2$ contains all the variances ordered exactly like in the case of $\mathbf{V}$.
Note that if all pixels share the same noise variance, equal to $\sigma_n^2$, the likelihood simplifies to
\begin{equation}
p(\mathbf{V}|\mathbf{Z}) = \mathcal{N}(\mathbf{V}|\mathbf{W} \mathbf{Z}, \sigma_n^2 \mathbf{I}).
\end{equation}

If we assume that all $Z(v)$ profiles are equally probable a-priori, the posterior
probability for $Z(v)$ is
\begin{equation}
p(\mathbf{Z}|\mathbf{V}) \propto \mathcal{N}(\mathbf{V}|\mathbf{W} \mathbf{Z}, \mathrm{diag}(\sigmabold^2))
\label{eq:posteriorLSD}
\end{equation}
where the posterior distribution is now seen as a distribution with respect to $\mathbf{Z}$. Eq. (\ref{eq:posteriorLSD}) 
represents the full probabilistic LSD solution. Given the special structure of the matrix $\mathbf{W}$, it is very simple to compute 
the maximum a-posteriori point estimate (the one that maximizes the posterior distribution) and one ends up
with the known LSD result of \cite{donati97} \citep[see also][]{semel09}:
\begin{equation}
Z_\mathrm{LSD}(v_j) = \frac{\sum_i \sigma_{ij}^{-2} \alpha_i V_i(v_j)}{\sum_i \sigma_{ij}^{-2} \alpha_i^2}.
\label{eq:LSD}
\end{equation}

\subsection{Bayesian hierarchical model}
From the Bayesian point of view, the classical LSD method is just the maximum a-posteriori solution to
the inference problem when flat priors are used for the common signature $Z(v)$. In other words, the value
of $Z(v)$ at one velocity position does not depend at all on the value at different velocity positions.
This assumption is not capturing some knowledge that we have from the physics of the problem. We know
that there must be some correlation between different velocity positions because the Stokes $V$ profile 
are a consequence of the combined solution of the radiative transfer equation and the statistical equilibrium
equations on the stellar atmosphere. The emission and absorption properties of the atmosphere inherently introduce
correlation between different velocity bins. One consequence of this is the typical smooth shape of the $Z(v)$
profile, that we should somehow force in our solution.

\begin{figure}
\centering
\includegraphics[width=\columnwidth]{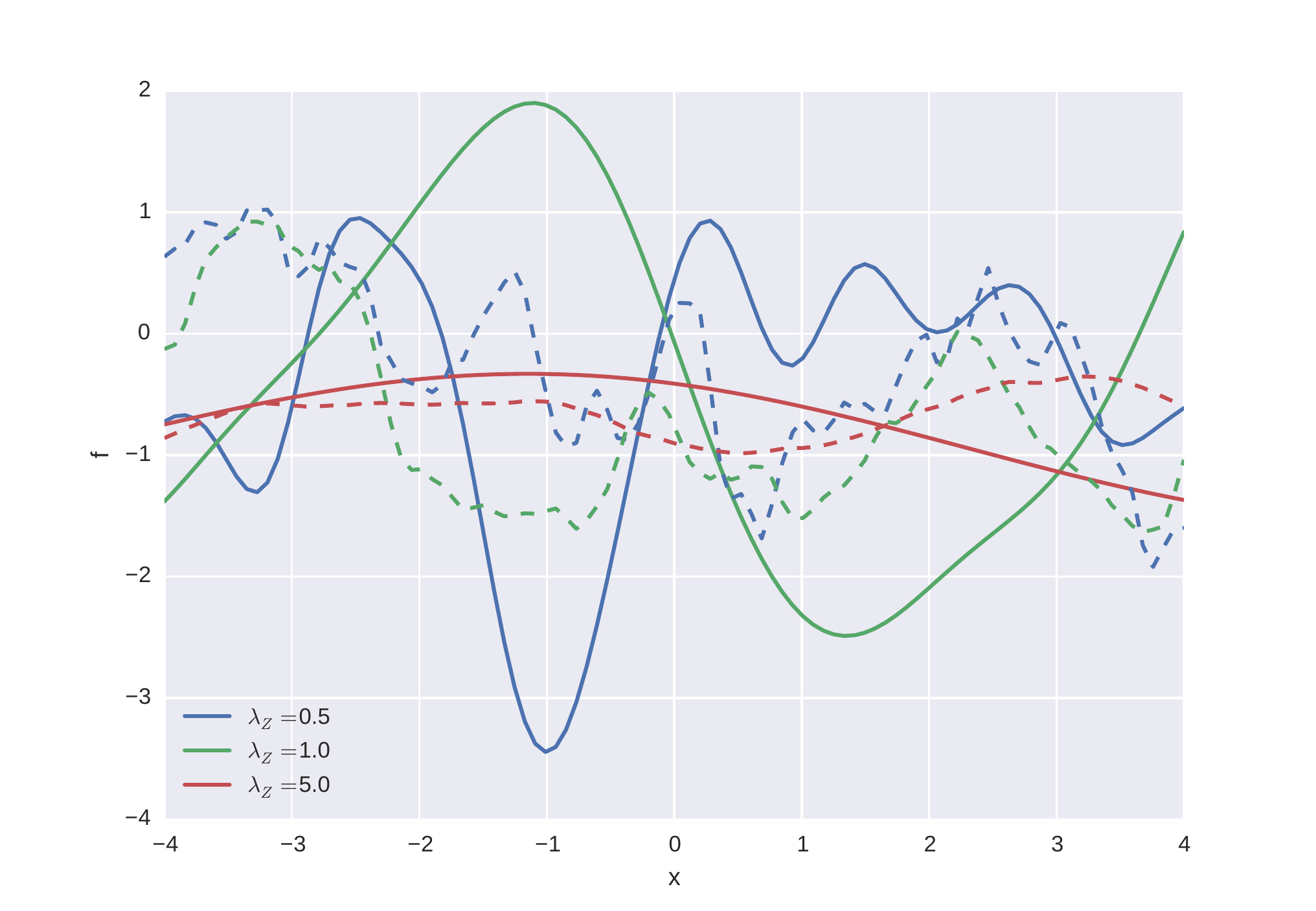}
\caption{Examples of functions extracted from GP priors with a squared exponential (solid lines) and Matern 
with $\nu=3/2$ covariance functions. In both cases, $\sigma_Z=1$ and different values of $\lambda_Z$ are 
used. Note that $\lambda_Z$ acts as a scale over which the function varies. The functions extracted from
the Matern covariance function have a rougher behavior than those of the squared exponential one.}
\label{fig:exampleGP}
\end{figure}

In the Bayesian framework, we impose this condition by introducing a suitable prior distribution. A very
convenient way to do it is by using a Gaussian Process \citep[GP;][]{rasmussenGP05}. Generally, a GP is a 
distribution over the subspace of functions that fulfill certain conditions. From the technical
point of view, once a function is sampled at a finite number of points, a GP for the function $Z(v)$
is defined by the following multivariate normal distribution:
\begin{equation}
p(\mathbf{Z}|\alphabold_Z) = \mathcal{N}(\mathbf{Z}|\mubold,\mathbf{K}(\alphabold_Z)),
\label{eq:priorGP}
\end{equation}
where $\mubold$ is the mean of the distribution, which is usually set to zero, and
$\mathbf{K}(\alphabold_Z)$ is the covariance matrix, that depends on a set of parameters $\alphabold_Z$.
GP are then defined by the specific covariance matrix. Instead of writing the full matrix, it is customary to 
define a function (the covariance function) that gives the value of the matrix elements. This 
covariance function has to fulfill certain conditions so that the ensuing covariance matrix
is proper (for instance, the covariance matrix has to be symmetric). Widespread covariance functions are the squared exponential:
\begin{equation}
K_{ij}(\sigma_Z,\lambda_Z) = \sigma_Z^2 \exp \left[ -\frac{(v_i-v_j)^2}{2 \lambda_Z^2}\right],
\end{equation}
or the Mat\'ern class of covariance functions:
\begin{equation}
K_{ij}(\sigma_Z,\lambda_Z) = \sigma_Z \frac{2^{1-\nu}}{\Gamma(\nu)} \left( \frac{\sqrt{2\nu}|v_i-v_j|}{\lambda_Z} \right)^{\nu}
K_\nu \left( \frac{\sqrt{2\nu}|v_i-v_j|}{\lambda_Z} \right),
\end{equation}
where $K_\nu(x)$ is the modified Bessel function \cite[see][]{abramowitz72} and $\nu$ is a real number controlling
the roughness of the resulting functions. Mat\'ern covariance functions converge to the squared 
exponential covariance function when $\nu \to \infty$.
Other covariance functions can be found in \cite{rasmussenGP05}. Fig. \ref{fig:exampleGP} shows
samples from a squared exponential GP prior with $\sigma_Z=1$ and different values of $\lambda_Z$. In the case of 
the squared exponential 
covariance function, one can see that the functions are smooth and the value of $\lambda_Z$ defines the rate of variability.

The two covariance functions described in the previous paragraph are stationary \citep{rasmussenGP05}, because
they only depend on the difference $v_i-v_j$, so that they are translation invariant. The experiments carried
out in Sections \ref{sec:syntheticExperiments} and \ref{sec:realExperiments} demonstrate that these covariance
functions do a good work. Although we do not pursue this objective here, it is possible to work with non-stationary covariance functions that 
are more localized and could possible explain better some features in some specific velocity bins \citep[similar to the work of][]{czekala15}.

A GP prior introduces the parameters $\alphabold_Z$ into the inference. These parameters of the prior are 
termed hyperparameters and have to be considered in the posterior distribution. This is the reason why the 
inference model is hierarchical. Therefore, the posterior distribution of the hierarchical model is now
given by
\begin{equation}
p(\mathbf{Z},\alphabold_Z|\mathbf{V}) = \frac{p(\mathbf{V}|\mathbf{Z}) 
p(\mathbf{Z}|\alphabold_Z) p(\alphabold_Z)}{p(\mathbf{V})},
\end{equation}
where we have added a hyperprior, $p(\alphabold_Z)$, for the hyperparameters.
Given that the hyperparameters are of no interest for obtaining the common signature,
they should be marginalized (integrated) from the posterior, so that we obtain the posterior
distribution for $\mathbf{Z}$:
\begin{equation}
p(\mathbf{Z}|\mathbf{V}) = \frac{1}{p(\mathbf{V})}\int \mathrm{d} \alphabold_Z p(\mathbf{V}|\mathbf{Z}) p(\mathbf{Z}|\alphabold_Z) p(\alphabold_Z).
\label{eq:posteriorZ}
\end{equation}
This integral cannot be computed in closed form in general. For this reason, we pursue here a Type-II maximum likelihood (ML) solution \cite{B96},
also known as empirical Bayes or maximum marginal likelihood. This method is used for the analytical
treatment of hierarchical models in which the posterior distribution is intractable. 
The strategy is to compute the marginal posterior for the hyperparameters:
\begin{equation}
p(\alphabold_Z|\mathbf{V}) = \frac{1}{p(\mathbf{V})} p(\alphabold_Z) \int d\mathbf{Z} p(\mathbf{V}|\mathbf{Z})
p(\mathbf{Z}|\alphabold_Z),
\end{equation}
and assume that this distribution is strongly peaked. Under this condition, one can obtain the set
of hyperparameters $\hat \alphabold_Z$ that maximize the marginal posterior $p(\alphabold_Z|\mathbf{V})$ and
use a Dirac delta function as prior distribution:
\begin{equation}
p(\alphabold_Z) = \delta(\alphabold_Z - \hat \alphabold_Z).
\end{equation}
In such a case, the integral in Eq. (\ref{eq:posteriorZ}) can be trivially carried out, so that we find 
\begin{equation}
p(\mathbf{Z}|\mathbf{V},\hat \alphabold_Z) = \frac{1}{p(\mathbf{V})} p(\mathbf{V}|\mathbf{Z}) p(\mathbf{Z}|\hat \alphabold_Z),
\end{equation}
where we make explicit in the posterior for $\mathbf{Z}$ that we are conditioning on a specific value of the hyperparameters.

The advantage of following this Type-II ML approach is that the solution becomes analytical. Making use of the
definition of the likelihood of Eq. (\ref{eq:likelihood}) and the prior of Eq. (\ref{eq:priorGP}) with $\mubold=0$ and
applying standard properties of the multivariate Gaussian distribution, we find:
\begin{equation}
p(\mathbf{Z}|\hat \alphabold_Z,\mathbf{V}) = \mathcal{N}(\mathbf{Z}|\mubold_Z,\Sigmabold_Z)
\label{eq:posteriorZNormal}
\end{equation}
where
\begin{align}
\Sigmabold_Z &= \left[ \mathbf{K}(\alphabold_Z)^{-1} + \mathbf{W}^T \mathrm{diag} \left( \frac{1}{\sigmabold^2} \right) \mathbf{W} \right]^{-1} \nonumber \\
\mubold_Z &= \Sigmabold_Z \mathbf{W}^T \mathrm{diag} \left( \frac{1}{\sigmabold^2} \right) \mathbf{V}
\label{eq:posteriorMeanCovZ}
\end{align}
Likewise, the marginal posterior for $\alphabold_Z$ is:
\begin{equation}
p(\alphabold_Z|\mathbf{V}) = p(\alphabold_Z) \mathcal{N}(\mathbf{V}|\mubold_\alpha,\Sigmabold_\alpha),
\label{eq:marginalAlpha}
\end{equation}
with
\begin{align}
\mubold_\alpha &= \mathbf{0} \nonumber \\
\Sigmabold_\alpha &= \mathrm{diag}(\sigmabold^2) + \mathbf{W} \mathbf{K}(\alphabold_Z) \mathbf{W}^T
\end{align}
The objective now is to obtain the value of $\hat \alphabold_Z$ that maximize Eq. (\ref{eq:marginalAlpha}). This is
easier if one takes logarithms, ending up with the maximization of 
\begin{equation}
\log p(\alphabold_Z|\mathbf{V}) = \log p(\alphabold_Z) -\frac{1}{2} \mathbf{V}^T \Sigmabold_\alpha^{-1} \mathbf{V}
- \frac{1}{2} \log \left| \Sigmabold_\alpha \right|,
\end{equation}
where $\left| \Sigmabold_\alpha \right|$ is the determinant of the matrix.

\begin{figure*}
\centering
\includegraphics[width=\columnwidth]{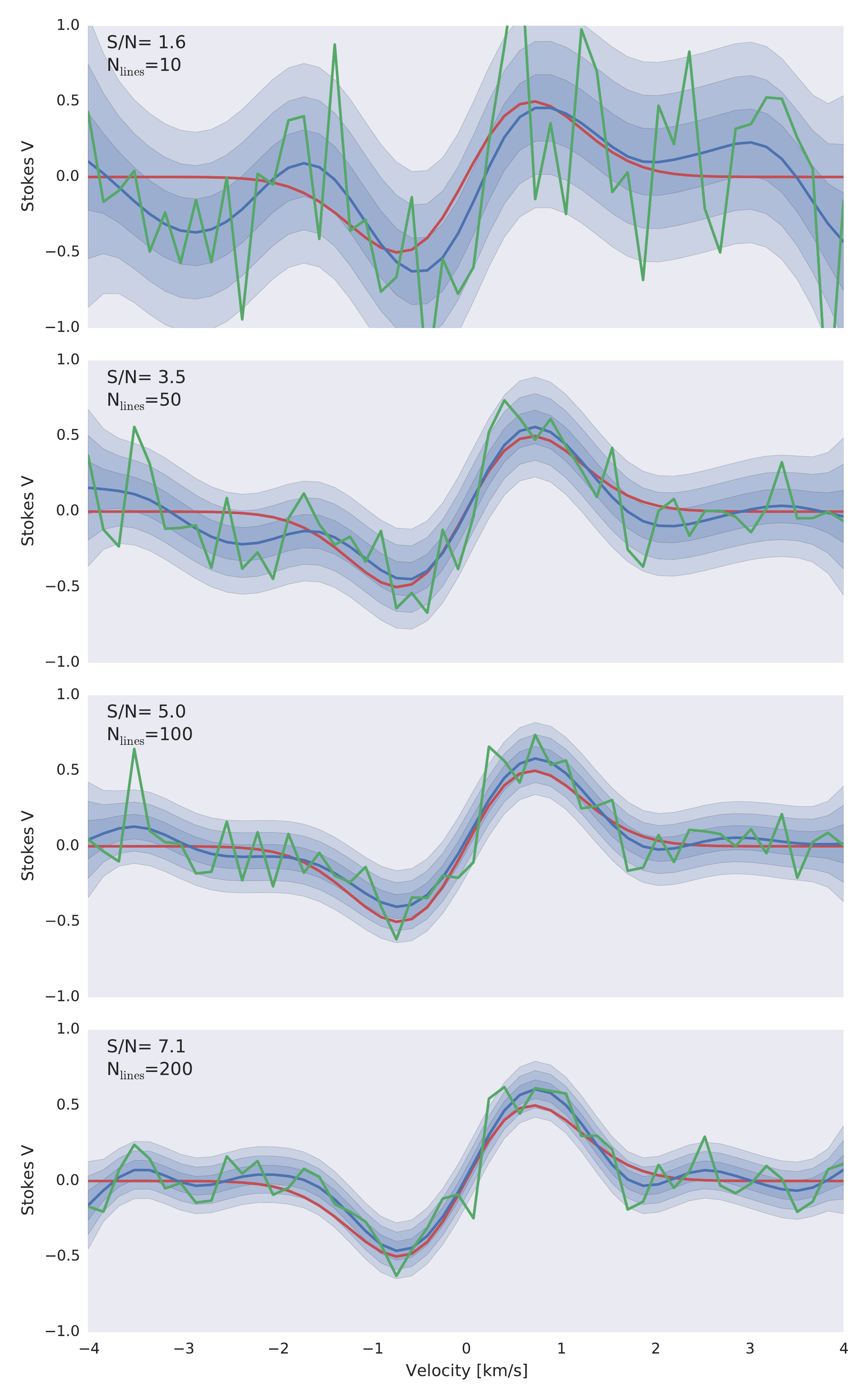}
\includegraphics[width=\columnwidth]{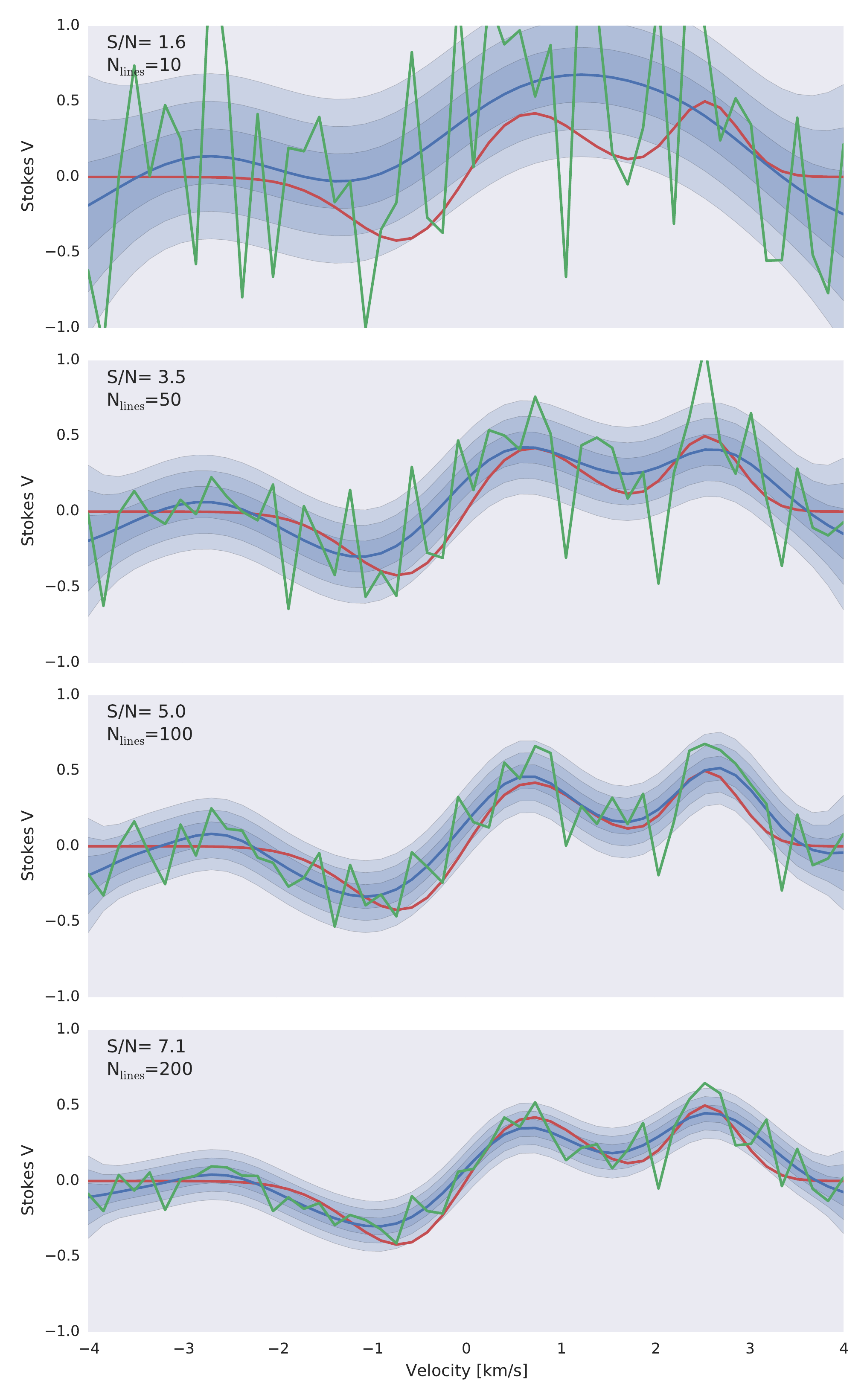}
\caption{Examples of two different shapes of $Z(v)$ and how they are recovered
using LSD (green curve) and Bayesian LSD (mean showed in blue and credibility intervals showed as blue shaded regions). The
original true $Z(v)$ is shown in red. The experiments use the Mat\'ern covariance function with $\nu=3/2$.}
\label{fig:exampleInference}
\end{figure*}

From a technical point of view, the calculation of the inverse and the determinant is computationally heavy. The reason is that
matrix $\Sigmabold_\alpha$ is of size $N_\mathrm{lines} N_v \times N_\mathrm{lines} N_v$, which is very large in practical
applications. However, it is possible to efficiently accelerate the computation using the Woodbury matrix identity \citep{matrixcookbook12},
so that
\begin{align}
\Sigmabold_\alpha^{-1} & = \mathrm{diag}\left( \frac{1}{\sigmabold^2} \right) \nonumber \\
&- \mathrm{diag}\left( \frac{1}{\sigmabold^2} \right) \mathbf{W} \left[ \mathbf{K}(\alphabold_Z)^{-1} + \mathbf{W}^T 
\mathrm{diag}\left( \frac{1}{\sigmabold^2} \right) \mathbf{W} \right]^{-1} \mathbf{W}^T \mathrm{diag}\left( \frac{1}{\sigmabold^2} \right).
\label{eq:inverseSigma}
\end{align}
In this case, it is only necessary to obtain the inverse of the $N_v \times N_v$ matrix $\mathbf{K}(\alphabold_Z)$ and of the
matrix inside brackets, which is again of size $N_v \times N_v$. These matrices are of a much smaller size than $\Sigmabold_\alpha$ and
the inversion is much less time consuming. Likewise, using the matrix determinant lemma, we find:
\begin{align}
\log |\Sigmabold_\alpha| &= \log \left| \mathbf{K}(\alphabold_Z)^{-1} + \mathbf{W}^T \mathrm{diag}\left( \frac{1}{\sigmabold^2} \right) \mathbf{W} \right| \nonumber \\
&+ \log \left| \mathbf{K}(\alphabold_Z) \right| + \log \left| \mathrm{diag}(\sigmabold^2) \right|.
\end{align}
The inverse of the matrix inside brackets can be efficiently computed using the Cholesky decomposition which also gives as the
logarithm of the determinant as a subproduct.

\begin{figure*}
\centering
\includegraphics[width=\columnwidth]{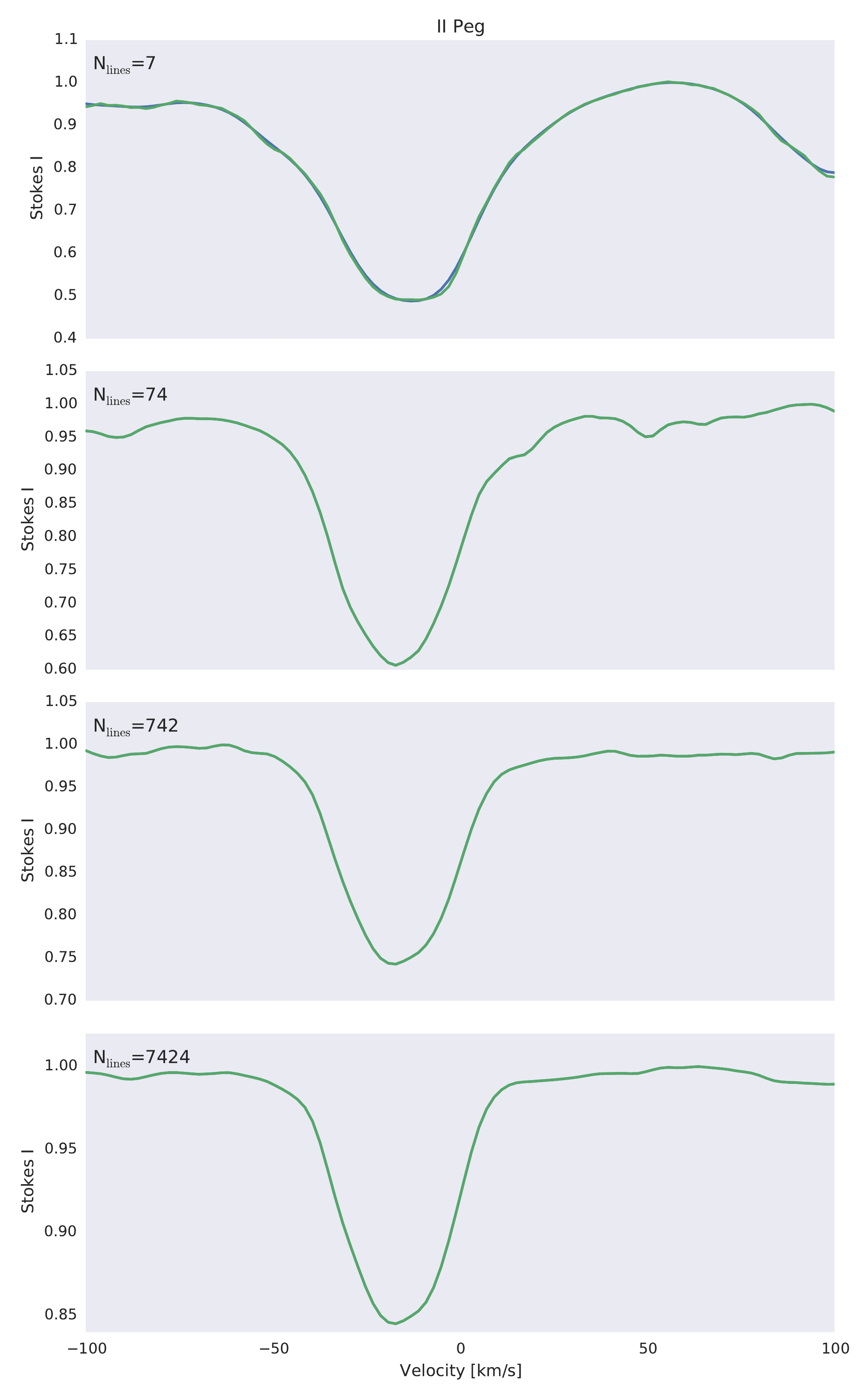}
\includegraphics[width=\columnwidth]{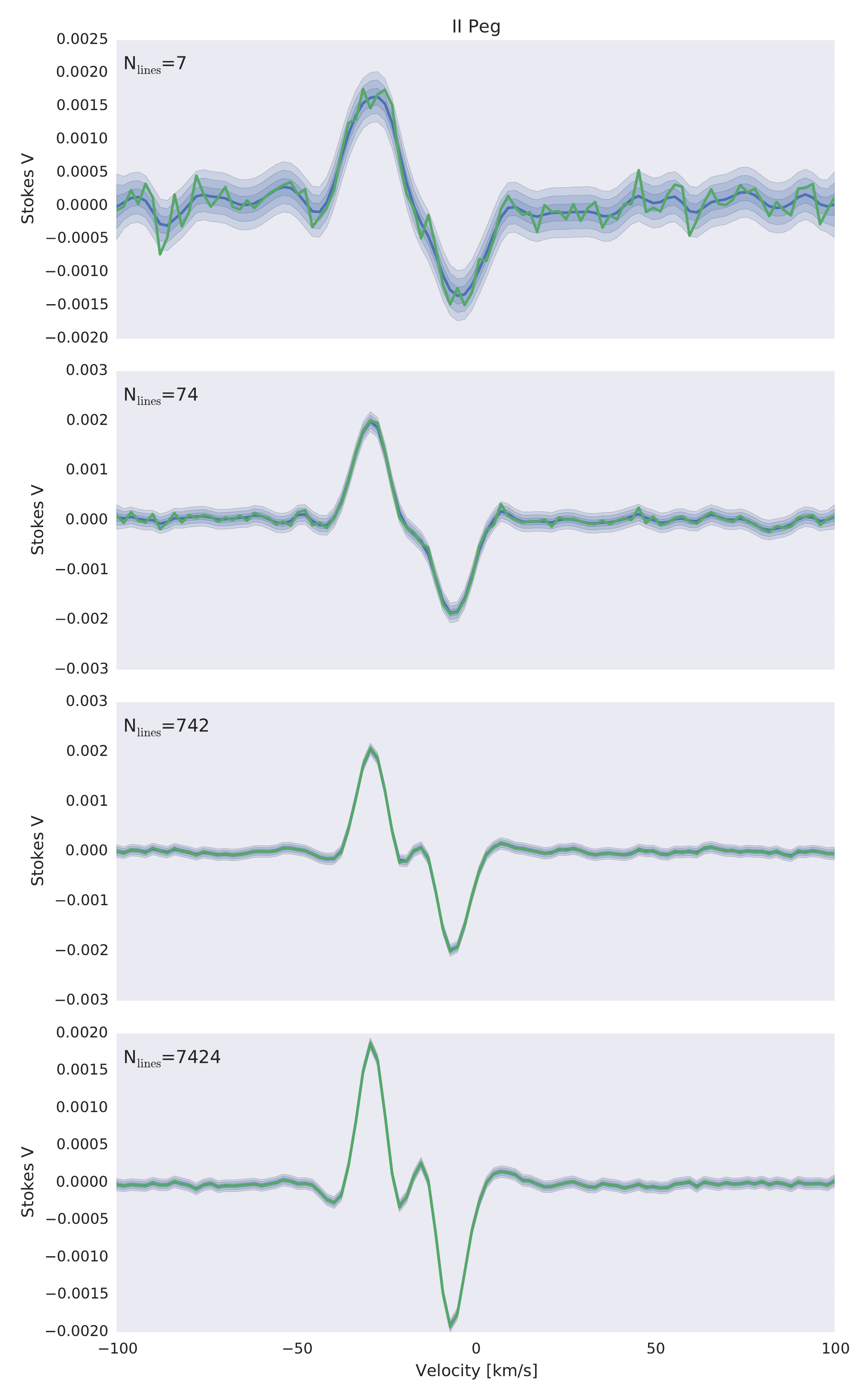}
\caption{Results of applying the LSD (green curve) and Bayesian LSD (blue curve showing the mean and blue regions showing the credibility intervals)
to observations of II Peg, both for Stokes $I$ (left panel) and $V$ (right panel). The experiments use the Mat\'ern covariance function with $\nu=3/2$.
The green and blue curves overlap almost exactly in the left panel when increasing the number of considered spectral lines.}
\label{fig:realInference1}
\end{figure*}

As a final remark, note that many of the terms in Eq. (\ref{eq:inverseSigma}) do not change during the optimization with 
respect to the hyperparameters. They can be precomputed at the beginning of the optimization and only the term
inside brackets in Eq. (\ref{eq:inverseSigma}) needs to be recomputed when the hyperparameters are changed.

\section{Synthetic experiments}
\label{sec:syntheticExperiments}
The first test of the method is done with synthetic data. We generate two artificial $Z(v)$ profiles, which are 
given by:
\begin{align}
Z_1(v) &= \exp \left[ -v^2 \right] \nonumber \\
Z_2(v) &= \exp \left[ -v^2 \right] + \frac{1}{5} \exp \left[ -(v-2.5)^2 \right].
\end{align}
For convenience, both profiles are normalized to $1/2$ amplitude, multiplied by randomly generated $\alpha$ coefficients (with zero mean and unit variance)
and then noise with diagonal covariance with variance $\sigma^2=1$ is added to all the synthetic observations. A maximum of 200 synthetic
spectral lines are used in both the standard LSD and the Bayesian LSD algorithms. For a single line, the signal-to-noise ratio equals one $S/N=1/2$.
It is expected that, in the presence of noise, adding $N_\mathrm{lines}$ spectral lines would increase the $S/N$ in proportion to $\sqrt{N_\mathrm{lines}}$.
Figure \ref{fig:exampleInference} shows how both estimations approach the true common signature (represented in red). The Bayesian
LSD result is using a Matern covariance function with $\nu=3/2$. The standard LSD
method gives the green estimation. It can be seen that, given that no correlation is imposed between velocity bins, a large
oscillation produced by noise remains even when adding 200 lines.

The mean of the marginal posterior for the common signature, given by $\mubold_Z$ in Eq. (\ref{eq:posteriorMeanCovZ}), is represented in blue. One of the
obvious advantages of working in a Bayesian framework is that the solution is given in terms of a probability distribution. This allows
us to compute the probability distribution for each velocity point. They are obtaining by marginalizing out all
variables in Eq. (\ref{eq:posteriorZNormal}) except for the value of $Z(v_j)$ at the velocity bin $j$ of interest. Using the special properties of the Gaussian distribution,
these distributions are just one-dimensional Gaussian distributions with variance equal to the $(j,j)$ element of matrix $\Sigmabold_{Z}$. 
We represent the distribution at each velocity point using the 68\%, 95\% and 99\% credibility intervals with blue shaded regions. 

\begin{figure*}
\centering
\includegraphics[width=\columnwidth]{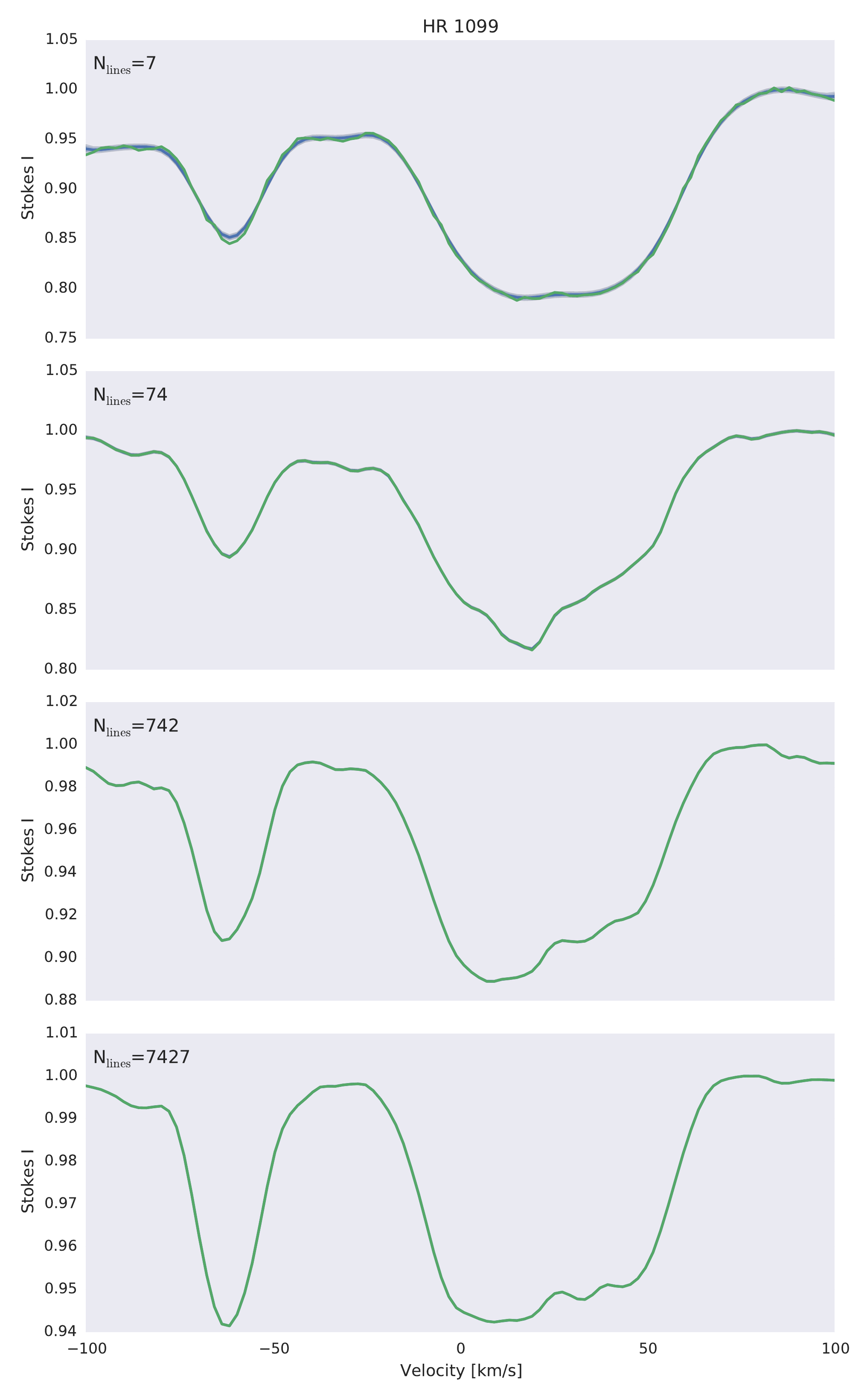}
\includegraphics[width=\columnwidth]{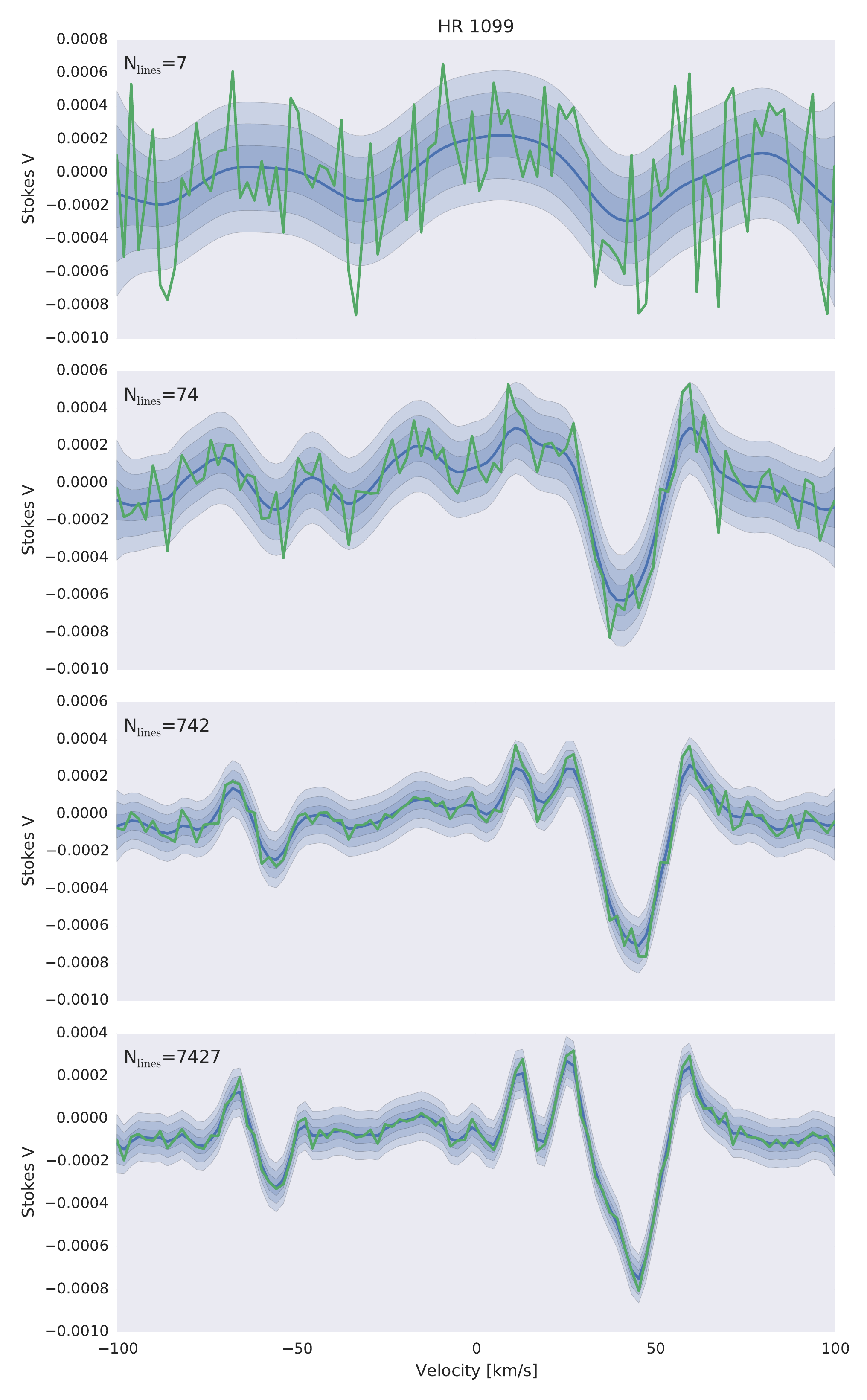}
\caption{Same as Fig. \ref{fig:realInference1} for HR1099.}
\label{fig:realInference2}
\end{figure*}

Several properties of the Bayesian LSD solution are worth mentioning. First, the mean of the solution is very smooth. This is a consequence of the
strong regularization introduced by the GP prior with squared exponential covariance function. Other covariance functions that give a less
smooth solution can also be used. Second, the covariance function is stationary (it only depends on $v-v'$) so that we find some variability
in the inferred solution when approaching the continuum. The reason is that the scale of variation $\lambda_Z$ is the same for all the
velocity bins. Many other covariance functions can be used to diminish this behavior. Third, the true solution is always inside the credibility 
intervals. Fourth, the marginal posterior of Eq. (\ref{eq:posteriorZ}) can be used as a suitable distribution to carry out the modeling of the
LSD profile. Using this distribution instead of the usual least squares solution (which assumes a diagonal covariance) makes the fitting much
more robust.

\section{Real data}
\label{sec:realExperiments}
The previous section demonstrates that Bayesian LSD can recover a much more robust
common signature from the observations, even when it is quite complex. In the following,
we apply the method to real observations. We consider three different cases, that are
representative of what one can encounter in typical observations. The case of II Peg is
representative of a very strong signal that can even be detected in individual lines. The
case of HR1099 is representative of a very complex common signature. Finally, 18 Sco is 
representative of a very weak signal in which, even taking into account thousands of lines,
the final signal is very close to the noise level. Public spectropolarimetric data for these three 
targets were downloaded from the PolarBase data base \citep{petit14}. 

\begin{figure*}
\centering
\includegraphics[width=\columnwidth]{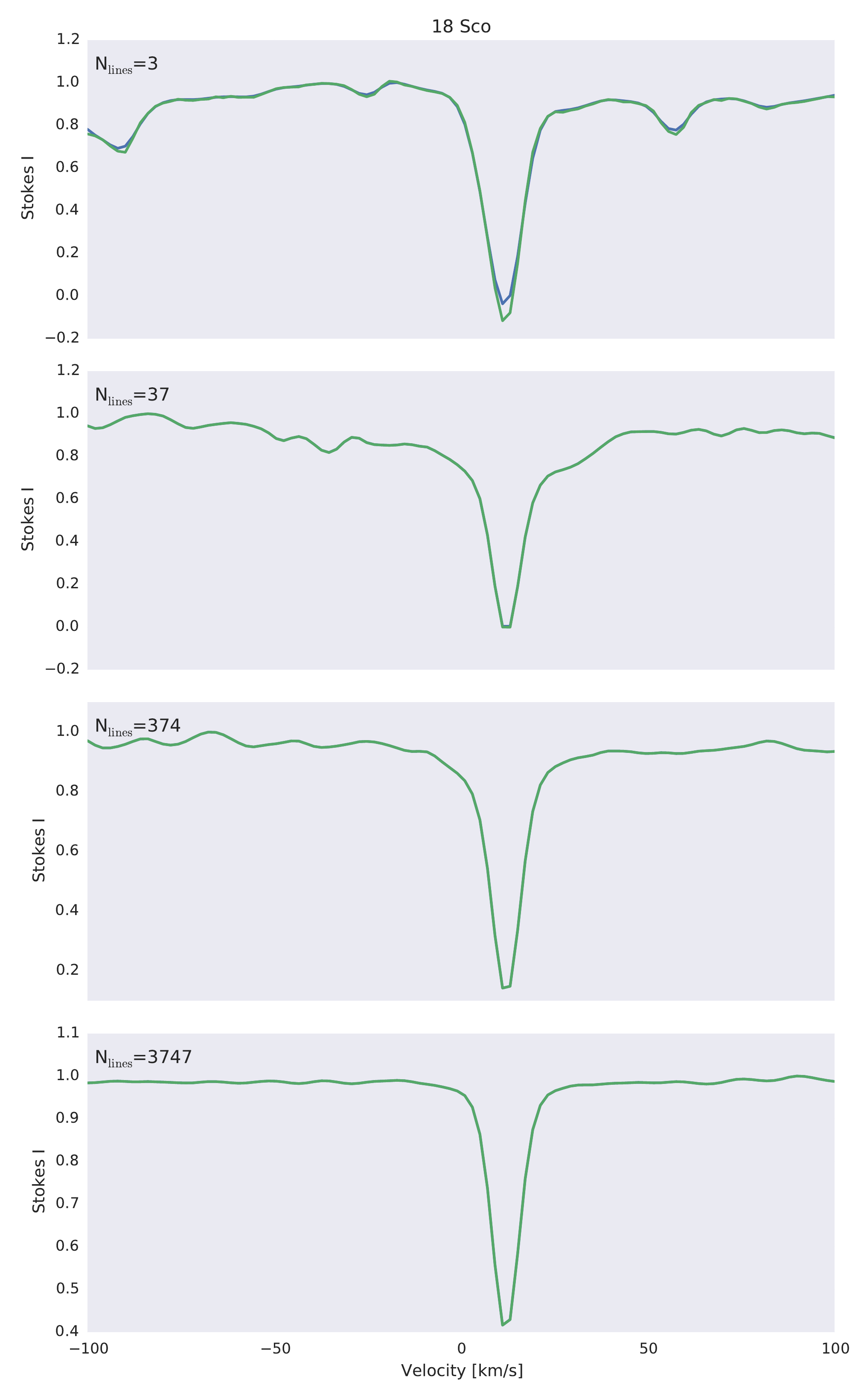}
\includegraphics[width=\columnwidth]{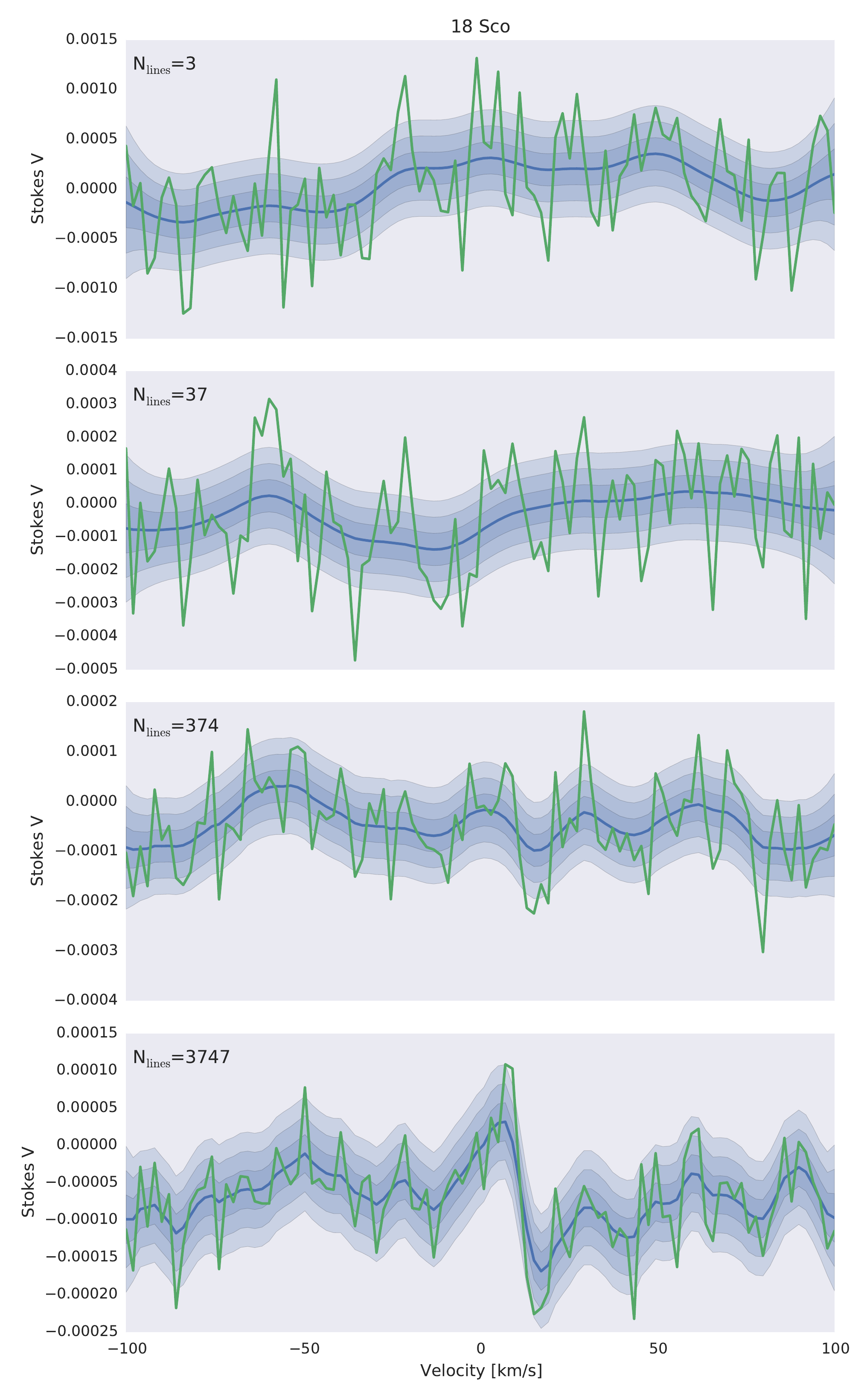}
\caption{Same as Fig. \ref{fig:realInference1} for 18 Sco.}
\label{fig:realInference3}
\end{figure*}

\subsection{II Peg}
II Peg is a very active RS CVn binary system, where a system of starspots has been mapped on the K1IV primary component.
The spot system usually covers up to 40\% of the surface \citep{berdyugina98}. The observations have
been carried out in July 2005 with the ESPaDOnS spectropolarimeter \citep{donati06}. 
Using a total integration time of one hour, the peak signal-to-noise ratio of the original 
Stokes $V$ spectrum (achieved at wavelengths around 780 nm) is close to 1,000 per wavelength bin. 
In agreement with the vast majority of published studies based on ESPaDOnS data, the 
$S/N$ used here is the one computed from the Stokes $I$ spectrum obtained by adding up the 
four polarimetric exposures combined to produce the Stokes V spectrum. 
For such extremely active target, this noise level is sufficiently low to detect Zeeman signatures in a number of 
individual spectral lines formed in the photosphere and chromosphere \citep{petit06}. Here, we use a list of about 7,400 spectral lines computed from an LTE atmospheric 
model matching the spectral type of the primary component of II Peg \citep{kurucz93b}, keeping only photospheric lines deeper than 40\% of the continuum level. 

Fig. \ref{fig:realInference1} shows the improvement in the detection of the common
signature $Z(v)$ when the number of lines that are considered increases. The left panel
shows the common signature in Stokes $I$, while the right panel displays the results
for Stokes $V$. The lines that we have considered have been 
ordered on decreasing value of the weights $\alpha_i$. Therefore, the panels with only 7 lines
are obtained considering the lines that produce the largest Zeeman signal. The LSD profile is shown
in green, while the mean of the Bayesian LSD detection is shown in blue, with the credibility intervals
at 68\%, 95\% and 99\% levels displayed as blue shades of different grades. For the case of II Peg, the signature is 
detected at a very high signal-to-noise ratio with only 7 lines (0.1\% of the total number of lines
considered in the line list). The Bayesian LSD approach leads to a much smoother profile than the LSD one.
The presence of credibility intervals is of relevance for discarding features like the one found at
$v \approx -85$km s$^{-1}$, which is compatible with zero in the Bayesian LSD case.
Increasing the number of spectral lines produce a convergence of the Bayesian LSD and the standard
LSD profiles, which is specially relevant in Stokes $V$. 

% Note that the residual uncertainty is \textbf{slightly} smaller inside the spectral line than in the
% continuum, something that is naturally captured by our Gaussian Process approach. The uncertainty
% in the continuum of Stokes $V$ remains still larger, giving the idea that signals below this detection limit
% cannot be distinguished.

\subsection{HR 1099}
Similarly to II Peg, HR 1099 is one of the most active RS CVn systems, with both components of
the binary association clearly visible in high resolution spectra (while the primary alone is seen in observations of II Peg).
The projected rotational velocity has been measured to be $v \sin i=41 \pm 0.5$ km s$^{-1}$ \citep{donati99}.
The magnetic structure of its active subgiant has been investigated in detail by \cite{petit04} using Zeeman Doppler
Imaging, showing the presence of large and axisymmetric regions with the magnetic field being
azimuthal. Additionally, \cite{petit04} also found clues for the presence of a differential
rotation, much weaker than in the case of the Sun. The observation considered here was taken with 
ESPaDONS in August 2005.

The line list used here is the same as the one previously employed for II Peg,
providing us with a good match to the photospheric characteristics of the evolved primary component, 
while the secondary is a much less active G5 main sequence star.

The presence of several magnetized regions that also change with time produce Zeeman signals
of large complexity. Figure \ref{fig:realInference2} shows the inferred
common signature for HR 1099, both for Stokes $I$ (left panel) and for Stokes $V$ (right panel). 
Several magnetic components are detected around $-65$ (which is roughly the 
radial velocity of the secondary star at that time), 20 and 40 km s$^{-1}$ (originating from the surface of the primary star of the system).
These detections in Stokes $V$ are also consistent with what is found in Stokes $I$.
When only the 7 most magnetically sensitive lines are considered, the mean of the Bayesian LSD profile for Stokes $V$ is very smooth, 
with the dashed regions marking the large uncertainty. A large part of the noise that is intrinsic to the 
standard LSD profile is absent. With only 74 lines, the feature at positive relative
velocities clearly appears. The quality of the detection increases drastically when
considering more than 10\% of the lines of the line list. However, it is interesting to
note that the uncertainty in the continuum regions does not decrease much when using 
10\% or 100\% of the lines (third and fourth panels). In the presence of only pure random
errors with constant variance, this uncertainty should be reduced roughly in proportion to $\sqrt{N_\mathrm{lines}}$.
However, the assumption of constant variance in these observations is not correct because the $S/N$ changes much 
throughout the ESPaDOnS spectrum, being much lower at the borders of individual spectral orders and also near the 
boundaries of the spectral domain of the instrument. This leads to a slower decrease of the final uncertainty.
It is important to point out that our Bayesian LSD method is able to take this into account and the ensuing final 
uncertainties do not decrease in such proportion. The immediate consequence of this is that we could be missing
very weak signals due to the presence of systematic errors.

\subsection{18 Sco}
18 Sco (HD 146233) is known to be one of the best solar twins among bright stars \citep{melendez07}.
In spite of its relatively low activity level, its surface magnetic field was detected by \cite{petit08} using high $S/N$ observations. The
Stokes $V$ spectrum analyzed here was obtained with the NARVAL spectropolarimeter \citep{auriere03} in July 2008.
This observation is extracted from the time-series presented by \cite{petit08}. Our line list is also identical to the one used 
in the same study. The LTE model from \cite{kurucz93b} was computed for a G2 spectral type, leading to a total
of about 3,700 photospheric spectral lines deeper than 40\% of the continuum.

The inferred common signature for 18 Sco is displayed in Fig. \ref{fig:realInference3} for
an increasing number of spectral lines. Even when the full line list is considered, only
a marginal detection of circular polarization is possible at the radial velocity of the star. This is an example in which having credibility intervals
for each velocity bin is crucial. The oscillations detected by the standard LSD method
fall inside the uncertainty region obtained from the Bayesian LSD method, although in this
case the mean is much smoother.

\section{Conclusions}
We have described a fully Bayesian least-squares deconvolution method which relies on a
Gaussian Process prior to regularize the spectral shape of the common signature. Two obvious 
advantages emerge from this Bayesian approach. First, the final answer to the problem is
in the form of a probability distribution. This allows the user to check whether
the detected signals are relevant or just a consequence of fluctuations of random and/or systematic
effects. Additionally, given that we also compute the covariance matrix of the ensuing probability 
distribution (taking into account possible correlations among all velocity bins), it can be used 
in inversion codes that fit the Bayesian LSD profile to estimate the stellar magnetic field.
This introduces a strong regularization that avoids fitting spurious effects. Examples of 
this are features of the LSD profiles that are close to the noise level or systematic effects
like interference fringes.
Second, the hierarchical approach using a GP prior automatically adapts to the 
complexity of the spectral profile. The solution for very noisy velocity bins (for instance,
close to the continuum) is very smooth and the noise is never overfitted. This automatic adaptation
to the signal is allowed due to the inclusion of the two hyperparameters of the GP prior. They are
easily inferred from the data.

By taking advantage of standard linear algebra identities and techniques for working with
sparse matrices, the Bayesian LSD common signature can be computed very fast in a matter of 
a few seconds for thousands of spectral lines. The code to compute 
the Bayesian LSD profile is freely available at \texttt{https://github.com/aasensio/pyGPLSD}.

\begin{acknowledgements}
Financial support by the Spanish Ministry of Economy and Competitiveness 
through projects AYA2010--18029 (Solar Magnetism and Astrophysical Spectropolarimetry) and Consolider-Ingenio 2010 CSD2009-00038 
are gratefully acknowledged. AAR also acknowledges financial support
through the Ram\'on y Cajal fellowships. 
This research has made use of NASA's
Astrophysics Data System Bibliographic Services.
\end{acknowledgements}

% \bibliographystyle{aa}
% \bibliography{/scratch/Dropbox/biblio}

\end{document}